\documentclass[a4paper,11pt]{article}
\pdfoutput=1 
\usepackage{jinstpub} 

\usepackage{subfigure}
\usepackage{textgreek}
\usepackage{lineno}

\title{Time Resolution Studies of Timepix3 Assemblies with Thin Silicon Pixel Sensors}

\author[a,b,1,2]{F.~Pitters\note{Corresponding author}\note{Now at HEPHY, Austria}}
\author[a]{N.~Alipour~Tehrani}
\author[a]{D.~Dannheim}
\author[a]{A.~Fiergolski}
\author[a,3]{D.~Hynds\note{Now at NIKHEF, The Netherlands}}
\author[a]{W.~Klempt}
\author[a]{X.~Llopart}
\author[a]{M.~Munker}
\author[a,4]{A.~N{\"u}rnberg\note{Now at KIT, Germany}}
\author[a]{S.~Spannagel}
\author[a,c]{M.~Williams}

\affiliation[a]{CERN, Switzerland}
\affiliation[b]{TU Wien, Austria}
\affiliation[c]{University of Glasgow, Great Britain}

\emailAdd{florian.pitters@pm.me}

\abstract{Timepix3 is a multi-purpose readout ASIC for hybrid pixel detectors. It can measure time and energy simultaneously by employing time-of-arrival (ToA) and time-over-threshold (ToT) techniques. Both methods are systematically affected by timewalk. In this paper, a method for pixel-by-pixel calibration of the time response is presented. Assemblies of Timepix3 ASICs bump-bonded to thin planar silicon pixel sensors with thicknesses of 50\,\textmu m, 100\,\textmu m and 150\,\textmu m are calibrated and characterised in particle beams. For minimum ionising particles, time resolutions down to 0.72\,$\pm$\,0.04\,ns are achieved.}


\begin{document}

\maketitle
\flushbottom


\section{Introduction}
\label{sec:introduction}

Timepix3~\cite{ref:timepix3} is a 65k channel multi-purpose readout ASIC from the Medipix family~\cite{ref:medipix,ref:timepix} for hybrid pixel detectors. Such detectors consist of a pixelated sensor material, silicon in this work, that is bonded to a readout ASIC with an equal pixel structure. By employing time-of-arrival (ToA) and time-over-threshold (ToT) techniques, each pixel in the Timepix3 ASIC can simultaneously measure the amplitude and time of signals from the sensor. However, both techniques display non-linear behaviour due to timewalk. To obtain the best possible performance, this effect has to be calibrated and corrected for on a pixel-by-pixel basis.

The performance of thin silicon sensors is of high interest for the CLIC vertex detector~\cite{ref:clic} which requires a spatial resolution of 3\,\textmu m and a time resolution of approximately 5\,ns at a material budget of 0.2\% radiation length per layer~\cite{ref:vertex}. While the Timepix3 geometry with 55\,\textmu m by 55\,\textmu m pixels will not enable a 3\,\textmu m spatial resolution, it is a suitable test vehicle to study thin sensors.

In this paper, the time resolution of the Timepix3 ASIC bonded to thin sensors is explored. A calibration method for pixel-by-pixel calibration of the ToA and ToT response by a combination of test pulse injection and beam data is presented. Timepix3 assemblies with silicon sensors of 50\,\textmu m, 100\,\textmu m and 150\,\textmu m thickness are calibrated and characterised in particle beams of minimum ionising particles (MIPs). Results on the time resolution obtained for different sensor thicknesses and bias voltages are presented.


\section{Experimental Setup and Methodology}
\label{sec:setup}
This section describes the details of ASIC and sensor as well as the further experimental setup and methodology that is used for the time measurements.

\subsection{The ASIC}
The Timepix3 ASIC comprises a matrix with 256 by 256 square pixels with 55\,\textmu m pitch. Each pixel consists of a charge sensitive amplifier (CSA) with a feedback capacitor for amplification and shaping of the input signal. If there is a voltage difference between the input and output node of the CSA, the capacitor is actively discharged via a current controlled by a Krummenacher feedback network~\cite{ref:krummenacher}. The discharge current is limited by the network, resulting in a constant discharge over time. The CSA output is connected to a single threshold discriminator with the possibility of a 4\,bit local threshold adjustment. The discriminator output is then connected to the in-pixel digital logic, where the signal is processed.

In the operating mode used in this work, the ASIC employs a 10\,bit ToT and a 14\,bit ToA counter running at a 40\,MHz clock for simultaneous measurements of amplitude and time. In this mode, the noise per pixel is approximately 90\,e-. The operating threshold is set to a value around 850\,e-. If the input signal rises above this value, the discriminator starts the ToT counter which then runs until the signal falls below the threshold again. The time the signal stays above threshold is a measure to the signal amplitude. At the time the signal crosses the threshold, the ToA value is taken from a global Gray code counter that is clocked at 40\,MHz. The ToA measurement is then refined by a local 4\,bit counter clocked at 640\,MHz which only runs until the next rising edge of the 40\,MHz clock. A more detailed description of the ASIC can be found in~\cite{ref:timepix3,ref:timepix3_tech}.

\begin{figure}[thb]
	\begin{center}
		\subfigure[]{
			\includegraphics[width=0.32\textwidth]{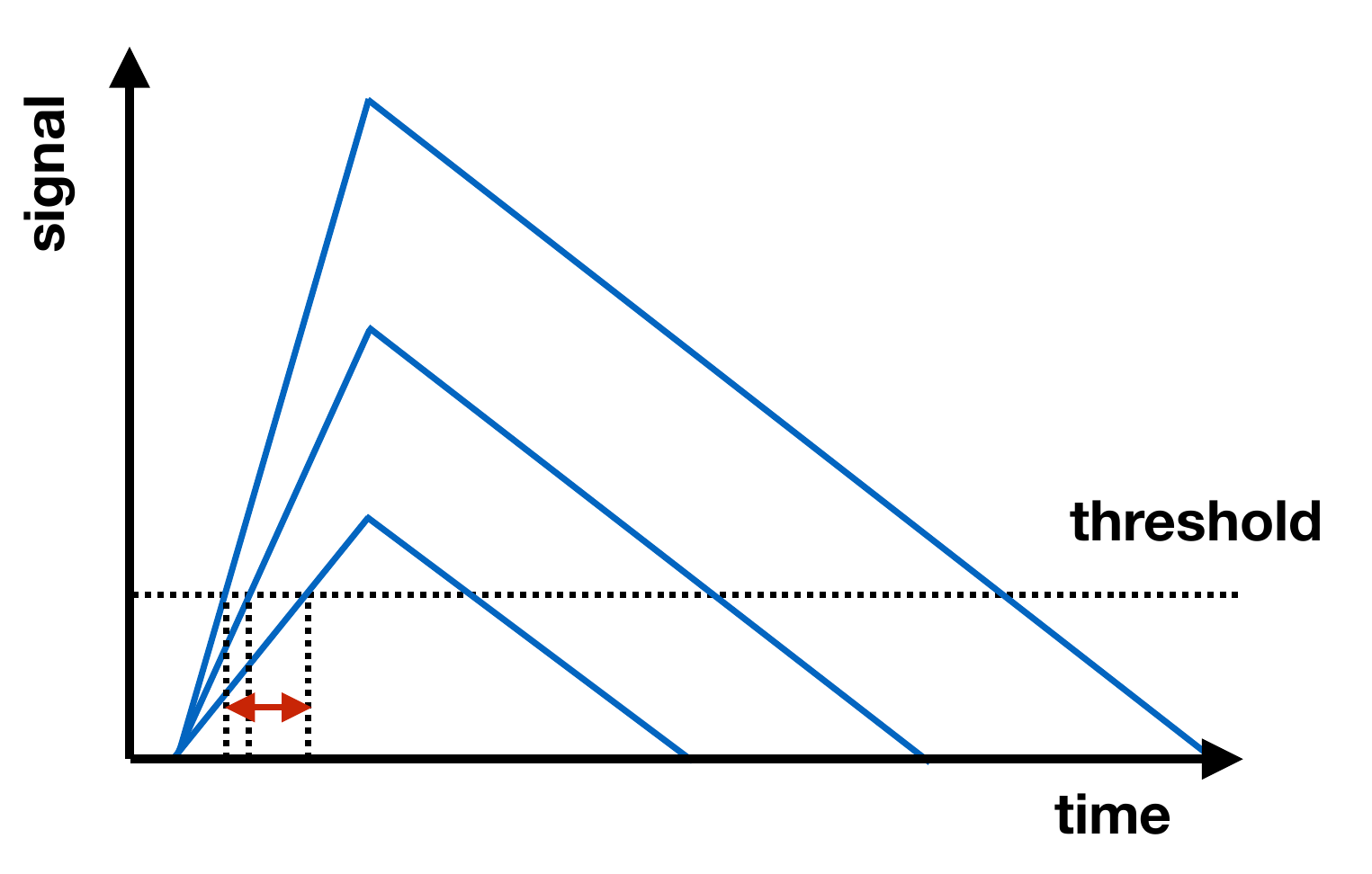}}%
		\subfigure[]{
			\includegraphics[width=0.32\textwidth]{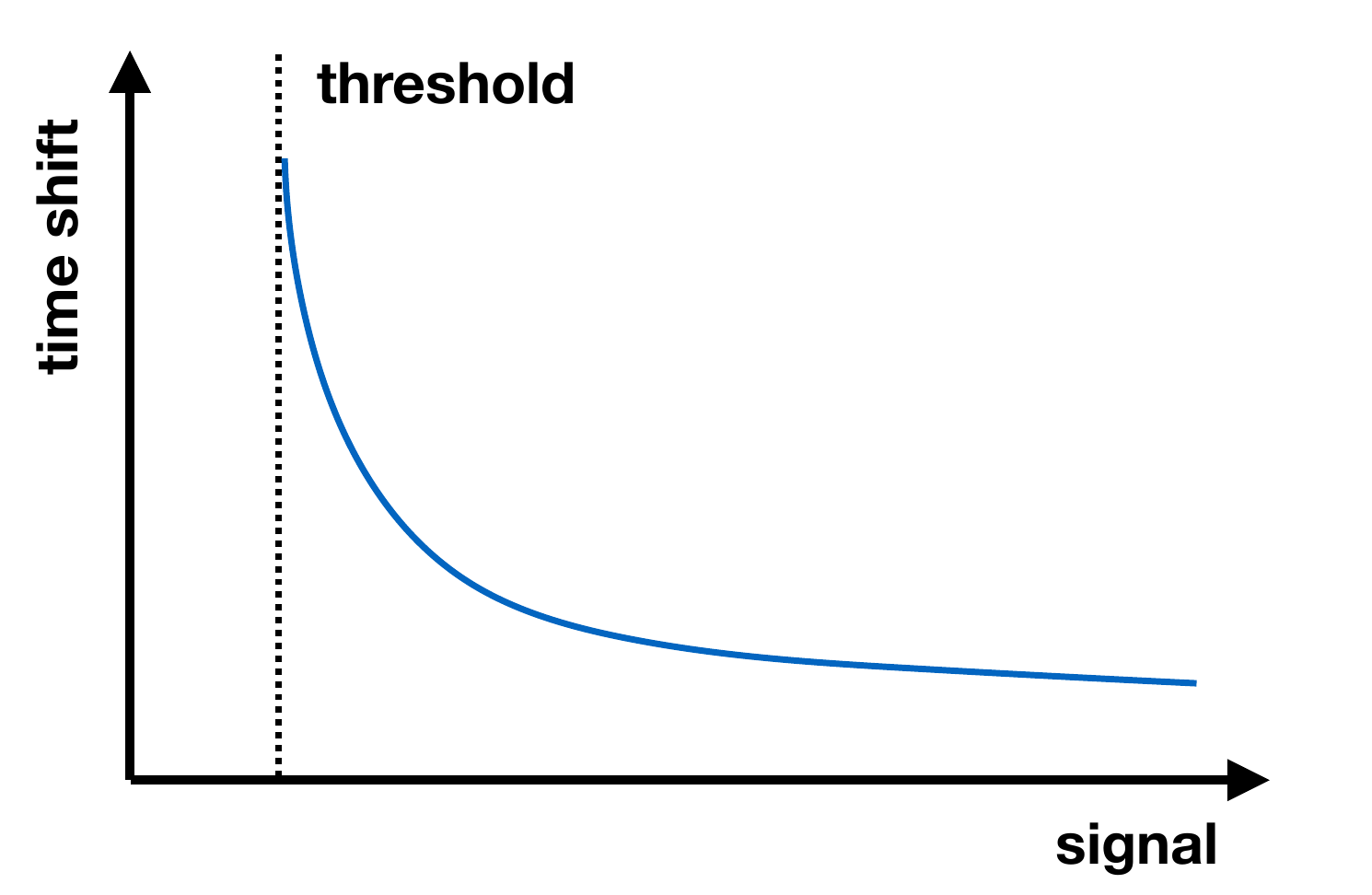}}%
		\subfigure[]{
			\includegraphics[width=0.32\textwidth]{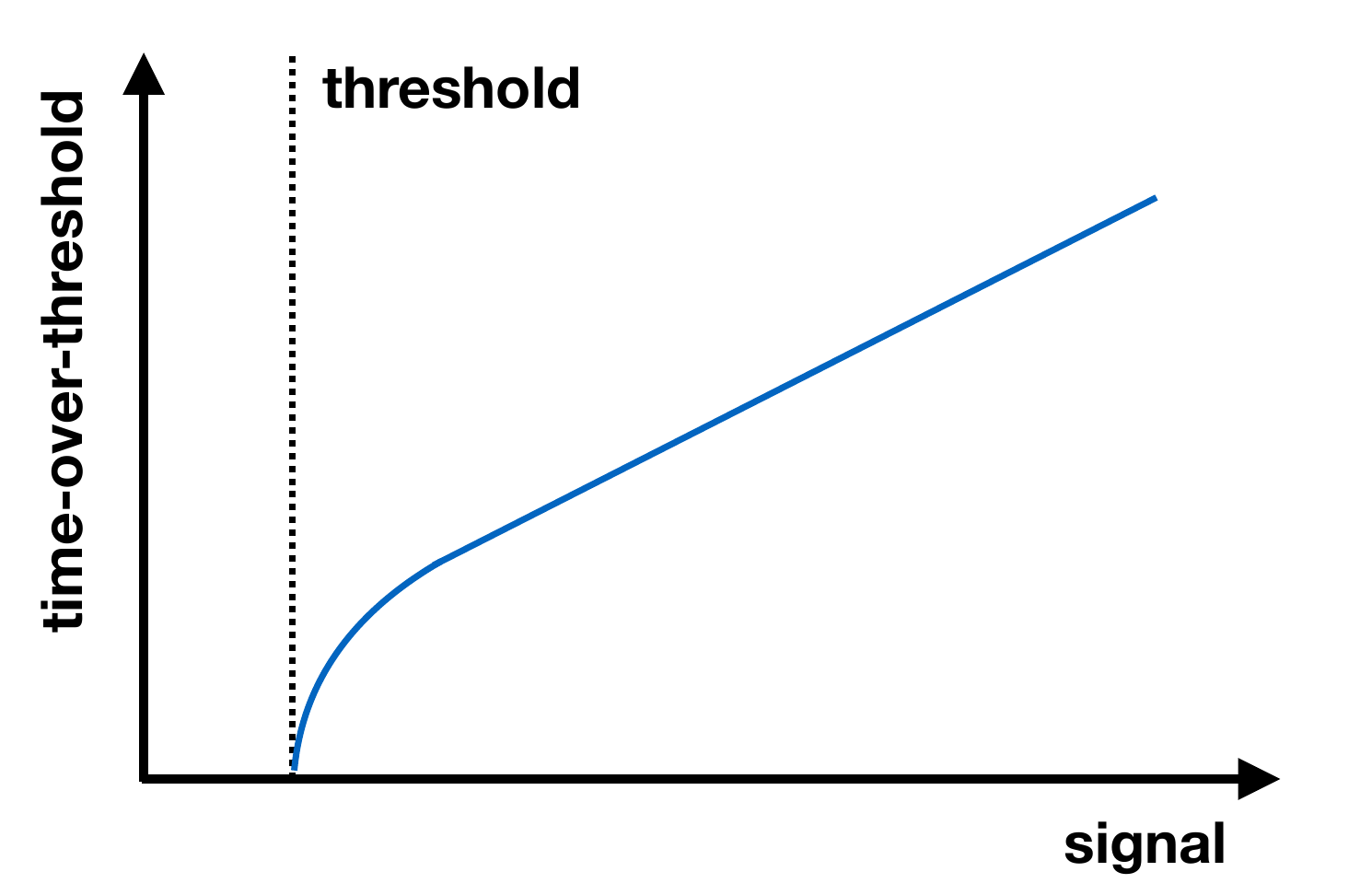}}
		\caption{Timewalk effect (a) resulting from signals with different amplitudes but constant rise time crossing a constant threshold on the ToA (b) and ToT (c) response.}
		\label{fig:tot_toa_schem}
	\end{center}
\end{figure}

Due to the use of a constant value threshold, the time at which the signal crosses the threshold depends on the signal amplitude. This phenomenon is commonly called timewalk. The shift in time $f_{\text{toa}}(x)$ for a signal with amplitude $x$ can be modelled with three free parameters b$_\text{toa}$, c$_\text{toa}$ and d$_\text{toa}$ as
\begin{equation}
	f_{\text{toa}}(x) = b_{\text{toa}} + \frac{c_{\text{toa}}}{x - d_{\text{toa}}}.
	\label{eq:toa}
\end{equation}
As the energy measurement is based on time as well, it suffers from the same phenomenon. Assuming a linear discharge, the energy response $f_{\text{tot}}(x)$ can be modelled with four free parameters a$_\text{tot}$, b$_\text{tot}$, c$_\text{tot}$ and d$_\text{tot}$ as
\begin{equation}
	f_{\text{tot}}(x) = a_{\text{tot}}\cdot x + b_{\text{tot}} - \frac{c_{\text{tot}}}{x - d_{\text{tot}}}.
	\label{eq:tot}
\end{equation}
Both models are based on simple geometrical considerations (see e.g. \cite[appendix A]{ref:pitters}) and are only approximate. For Timepix3 however, the work quite well. The effect of timewalk on both ToA and ToT is sketched in Fig.~\ref{fig:tot_toa_schem}.

Each pixel of the ASIC has the possibility for test pulse injection. Here, the test charge is injected by applying a step voltage to a 3\,fF capacitor that is connected in parallel to the sensor input. The injection input can be activated and deactivated separately for each pixel. An ADC in the chip periphery allows the measurement of the applied voltage amplitude. There is also the option to inject a test pulse directly into the in-pixel digital logic, bypassing the preamplifier and discriminator.

\subsection{The Sensors}
The Timepix3 is bump-bonded to planar active-edge n-on-p silicon pixel sensors. Thicknesses of 50\,\textmu m, 100\,\textmu m and 150\,\textmu m thickness are investigated. They are fully depleted at approximately 10\,V, 15\,V and 25\,V, respectively. A list of sensors and operating voltages used in this work is given in Tab.~\ref{tab:assemblies}.
Further results of these sensors with emphasis on the active-edge behaviour can be found in \cite{ref:sensors}.

\begin{table}
\begin{small}
\begin{center}
  \caption{List of assemblies and operating conditions for which results are shown in this work.}
  \begin{tabular}{cccccc} \\
    \hline\hline \\[-0.3em]
    \textsc{ID} & \textsc{Sensor Thickness} & \textsc{Full Depletion Voltage} & \textsc{Operating Voltages} \\
		\, & [\textmu m] & [V] & [V] \\[0.3em]
    \hline \\[-0.8em]
    1 & 50 & 10 & 10, 15, 20, 30, 40, 50 \\
    2 & 100 & 15 & 20 \\
    3 & 150 & 25 & 10, 15, 20, 30, 40, 50, 60 \\[0.7em]
    \hline\hline
  \end{tabular}
\label{tab:assemblies}
\end{center}
\end{small}
\end{table}

\subsection{Data Acquisition}
\begin{figure}[thbp]
	\begin{center}
			\includegraphics[height=0.32\textheight]{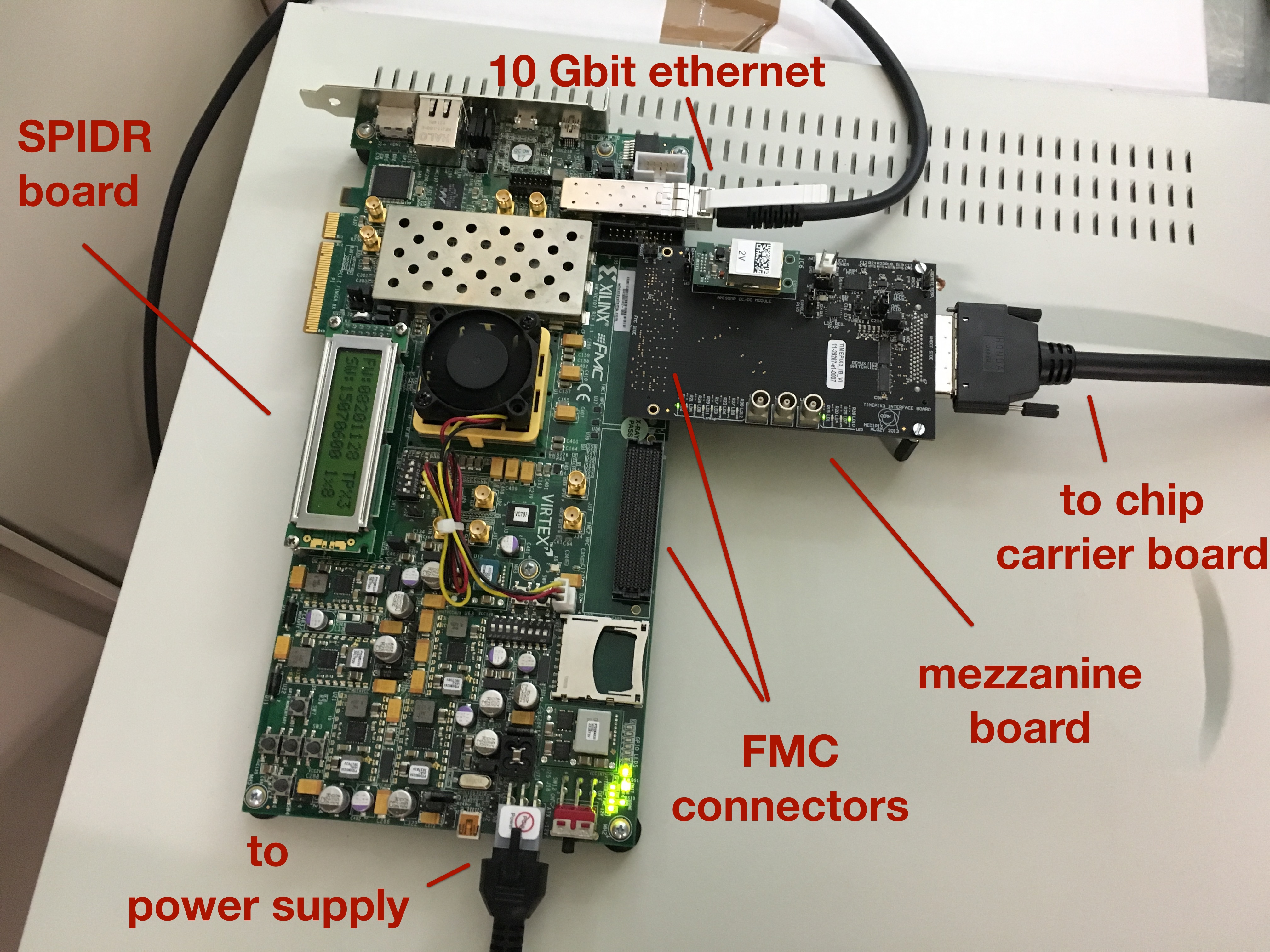}
		\caption{Close up of the SPIDR readout system. The system is described in detail in~\cite{ref:spidr}.}
		\label{fig:setup}
	\end{center}
\end{figure}

The data acquisition (DAQ) system is shown in Fig.~\ref{fig:setup}. The assembly of ASIC and sensor is wire-bonded to a so called chip carrier board, which handles mainly the power distribution for the ASIC and the bias voltage supply for the sensor (not shown in the picture). The chip carrier board is then connected to a SPIDR readout board, either directly or via a mezzanine board. There, an FPGA manages the data acquisition and slow control. For more information on the SPIDR data acquisition system, the reader is referred to~\cite{ref:spidr}. One SPIDR board can hold two chip carrier boards. A time-to-digital converter (TDC) channel with a precision of about 250\,ps is available for an external trigger input. The recorded data is transmitted via a 10\,Gbit Ethernet connection to a PC.

\subsection{Testbeam Setup and Reconstruction}
\label{sec:setup_testbeam}
\begin{figure}[thbp]
	\begin{center}
		\includegraphics[width=0.99\textwidth]{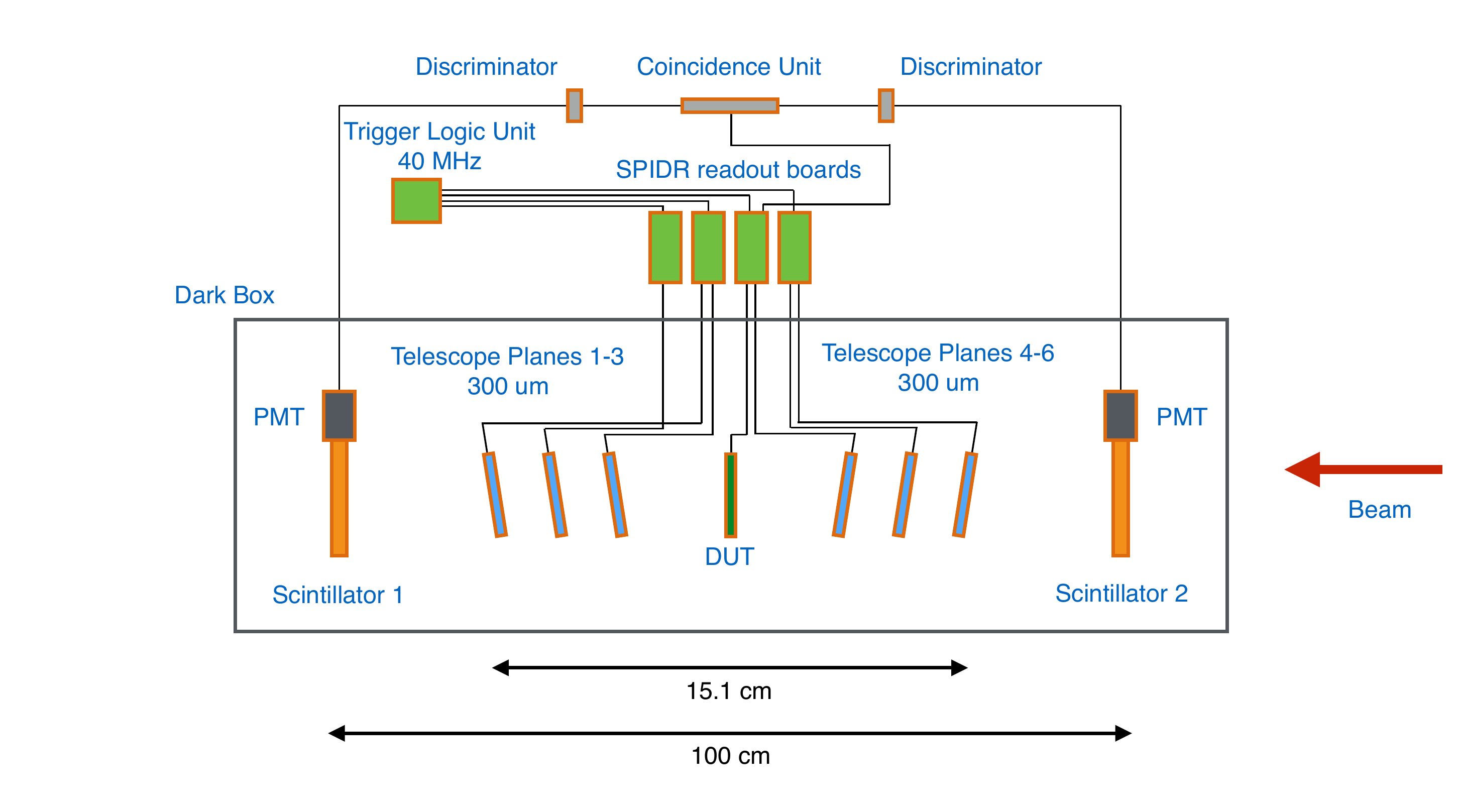}
		\caption{Schematic of the telescope setup and readout used for beam tests.}
		\label{fig:tb_setup}
	\end{center}
\end{figure}

Beam tests have been conducted at CERN in the SPS H6 beamline with 120\,GeV/c incident pions. The Timepix3 assemblies have been measured in the CLICdp telescope which is shown in Fig.~\ref{fig:tb_setup}. The telescope employs six tracking planes of Timepix3 assemblies, each with a 300\,\textmu m thick sensors. Three planes are upstream and three are downstream of a device under test (DUT). In total, four SPIDR systems are used to read the seven assemblies. The telescope planes are tilted by 9 degrees to optimise charge sharing. This way, a pointing resolution at the DUT position of about 1.8\,\textmu m~\cite{ref:nilou} is achieved. The track time resolution is about 1.1\,ns~\cite{ref:pitters}. Two organic scintillators read out by photomultiplier tubes (PMT) in the front and in the rear can be used to obtain a more precise reference timestamp. The telescope itself is not triggered but all recorded hits are read out and matched offline via their timestamps. Hits on the scintillators are only read out if they are recorded in coincidence with each other. They are then sampled by the TDC input of the second SPIDR board.

For the track reconstruction, the data stream is divided into slices of 25\,\textmu s. Clusters on the six planes that coincide within $\pm$200\,ns and $\pm$200\,\textmu m to each other are associated with a track candidate and reconstructed with a straight line fit. If the fit passes a \textchi$^2$/ndof criterion of less than 12/4 it is considered a track. On average, eight tracks are found within the 25\,\textmu s.

Three timestamps can then be extracted from track, scintillators and DUT. The track timestamp is taken as the average of the six associated clusters on the six telescope planes, corrected for average time-of-flight and latencies. Per cluster, the timestamp of the pixel with the largest energy deposit is assigned as cluster timestamp. The DUT timestamp is obtained in the same way. Lastly, a coincidence scintillator timestamp is associated with a track if one can be found within a window of $\pm$7\,ns around the track timestamp.

\begin{figure}[thbp]
	\begin{center}
		\includegraphics[width=0.80\textwidth]{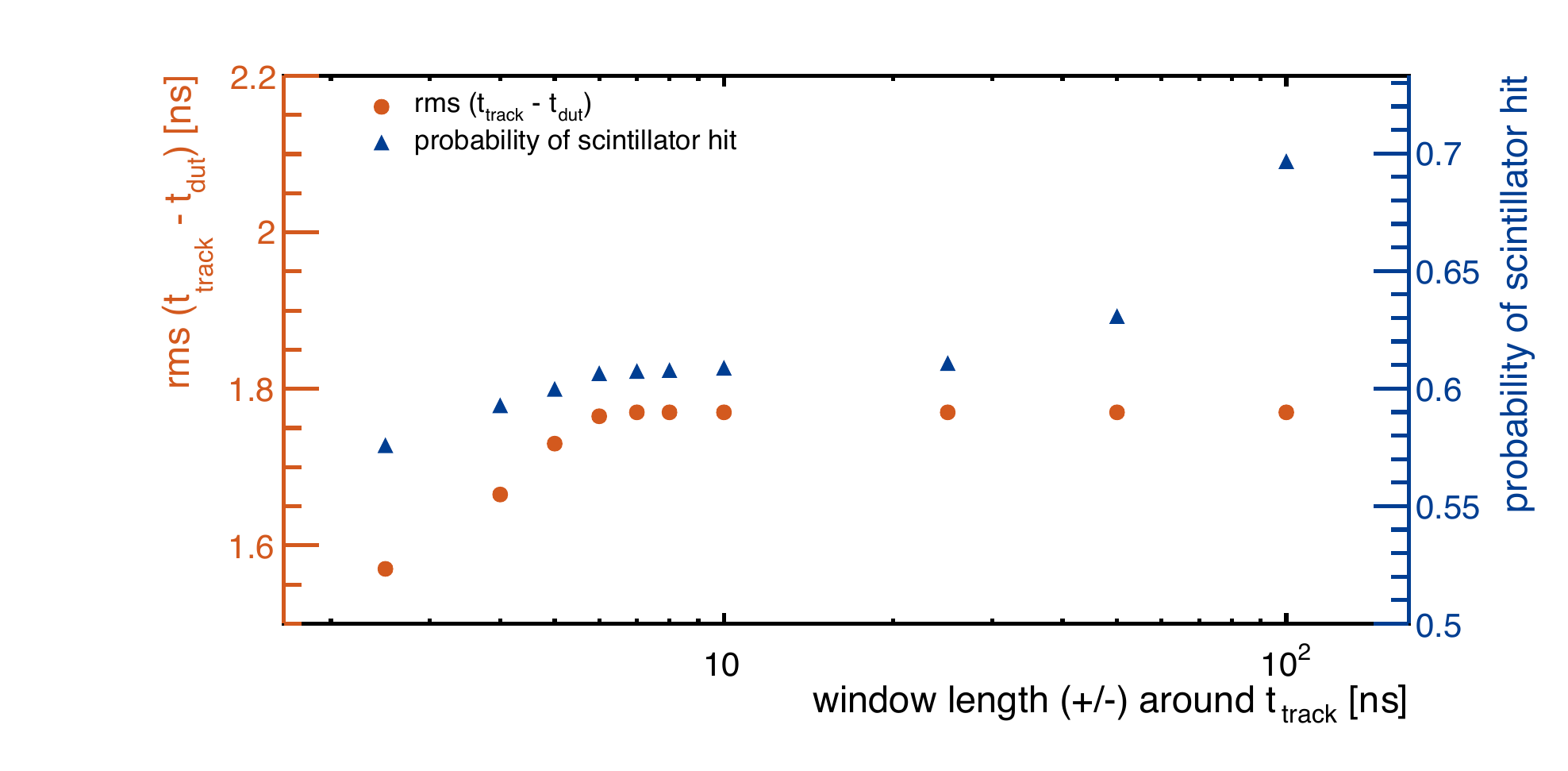}
		\caption{Influence of the acceptance window around t$_\text{track}$ used to search for a scintillator timestamp on the measured RMS between track and DUT timestamps.}
		\label{fig:trigger_window}
	\end{center}
\end{figure}

The scintillator has a better time resolution than the telescope tracks and is therefore the preferred time reference. However, due to the low efficiency of the scintillators (about 75\% per device) and the untriggered data stream, a finite window length is necessary to avoid association with a wrong track. The value of $\pm$7\,ns is motivated by Fig.~\ref{fig:trigger_window} which shows the RMS of the time difference between track and DUT for different window lengths as well as the probability of finding a coincident scintillator hit within that window. Since the three timestamps are all independent, the length of the acceptance window used to search for a scintillator timestamp must not bias the measured RMS between track and DUT timestamps. A window of $\pm$7\,ns is the smallest value that fulfils this criterion. The probability of finding a coincident scintillator hit within this window is 61\%, in agreement with previous lab measurements. At smaller windows, the probability and RMS of the residual t$_\text{track}$\,-\,t$_\text{dut}$ distribution starts to drop as the distribution becomes biased towards smaller time differences. At larger windows, the probability of finding a scintillator hit rises due to mis-association. The coincident scintillator timestamp is then used as a time reference and tracks are used only as position reference. Tracks without associated scintillator hits are discarded from the analysis.

\subsection{Time Measurement}
\label{sec:time_measurement}
From the three obtained timestamps of track, DUT and scintillator, three time differences can be calculated per track. The RMS values of the corresponding histograms then depend on the individual time resolutions $\sigma$.
\begin{align}
		\text{RMS}_{t_\text{dut}-t_\text{track}} &= \sqrt{\sigma_\text{dut}^2 + \sigma_\text{track}^2 } \\
		\text{RMS}_{t_\text{dut}-t_\text{ref}} &= \sqrt{\sigma_\text{dut}^2 + \sigma_\text{ref}^2 }  \\
		\text{RMS}_{t_\text{ref}-t_\text{track}} &= \sqrt{\sigma_\text{ref}^2 + \sigma_\text{track}^2 }
\end{align}
Assuming they are not correlated, the time resolution of the scintillator $\sigma_\text{ref}$ can be extracted.
\begin{equation}
		\sigma_\text{ref} = \sqrt{ -\frac{1}{2}(\text{RMS}_{t_\text{dut}-t_\text{track}}^2 - \text{RMS}_{t_\text{dut}-t_\text{ref}}^2 - \text{RMS}_{t_\text{ref}-t_\text{track}}^2)}
		\label{eq:ref_time}
\end{equation}
Indeed, the three time differences in Eq.~\ref{eq:ref_time} are not completely independent because all three time measurements depend on the same clock supplied by the Trigger Logic Unit (TLU). However, due to the much finer binning of the TDC that measures the scintillator as well as the combination of multiple timestamps for the track, each with a statistically independent component from time-of-flight and the charge carriers drift within the silicon, the correlation is assumed to be negligible.\footnote{Unfortunately, the (non) correlation can not easily be quantified as sensor effects are involved. To measure it, an independent time measurement with higher precision would be required.} Eq.~\ref{eq:ref_time} is applied independently to 25 analysed runs, ranging over different sensor thicknesses, bias voltages and calibration procedures, and the average extracted resolution of the scintillator is
\[ \sigma_\text{ref} = 0.30 \pm 0.04\text{ ns}. \]
The DUT resolution is defined as the RMS of the residual distribution between scintillator and DUT, with the scintillator resolution subtracted in quadrature.
\begin{equation}
	\sigma_\text{dut} = \sqrt{ \text{RMS}_{t_\text{ref} - t_\text{dut}}^2 - \sigma_\text{ref}^2 }
\end{equation}
The uncertainty on the RMS is estimated as 0.03\,ns from three different 50\,\textmu m assemblies measured at 15\,V. The propagated uncertainty on the time resolution $\sigma_\text{dut}$ is 0.04\,ns.


\section{Calibration of the Time Response}
\label{sec:calibration}
To reach the best possible timing performance, each pixel has to be calibrated independently. Due to the non-linear behaviour of the timewalk and the large number of pixels, this is done with a combination of test pulses and beam data. With this approach, only about 100 hits from minimum ionising particles per pixel are required for a full calibration. An alternative approach, using only beam data, is likely to require about two orders of magnitude more hits per pixel to reach the same precision in the non-linear part.

\subsection{Timewalk Calibration}
\label{sec:tw_calibration}
\begin{figure}[thbp]
	\begin{center}
		\subfigure[]{
			\includegraphics[width=0.49\textwidth]{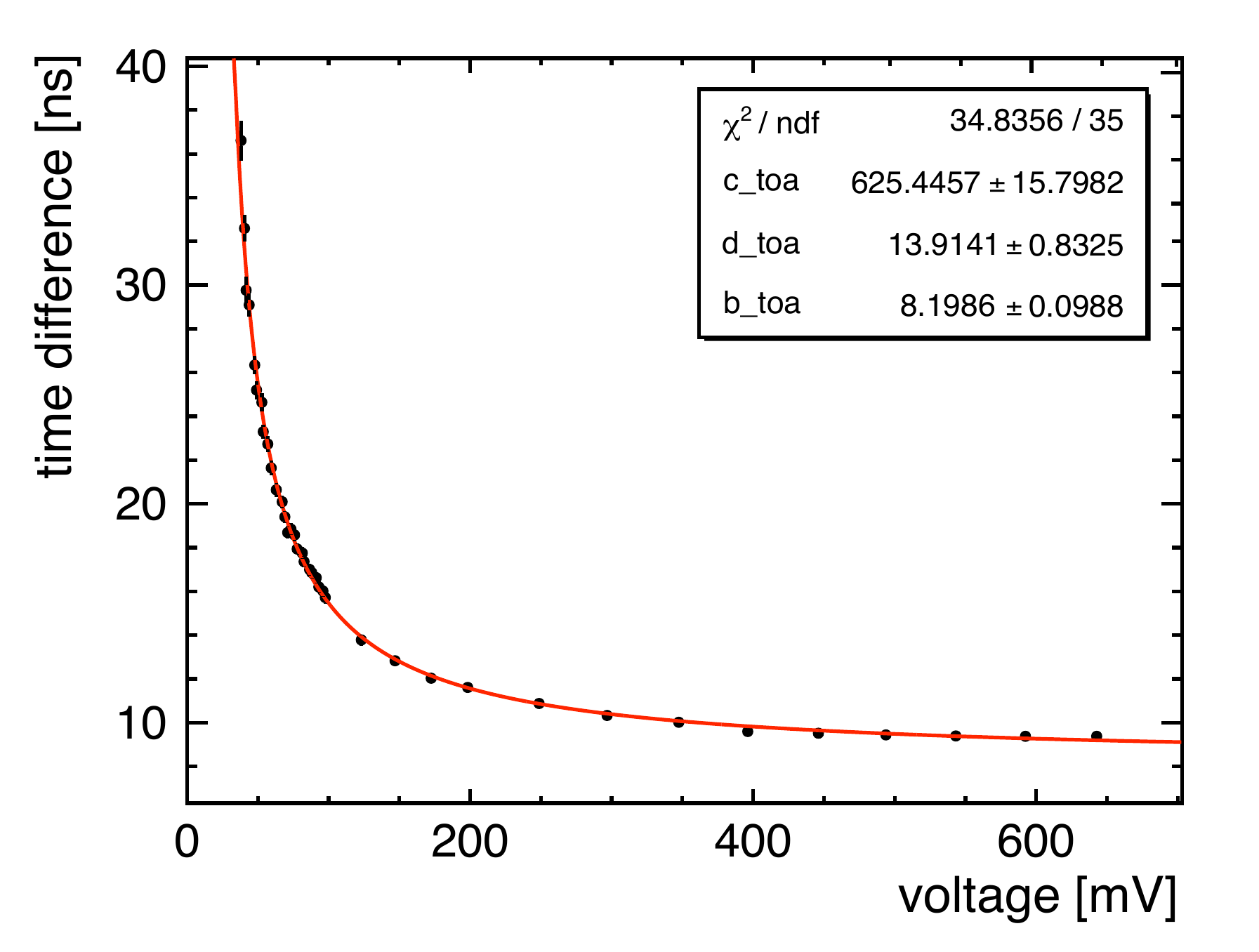}
			\label{fig:toa_example}}%
		\subfigure[]{
			\includegraphics[width=0.49\textwidth]{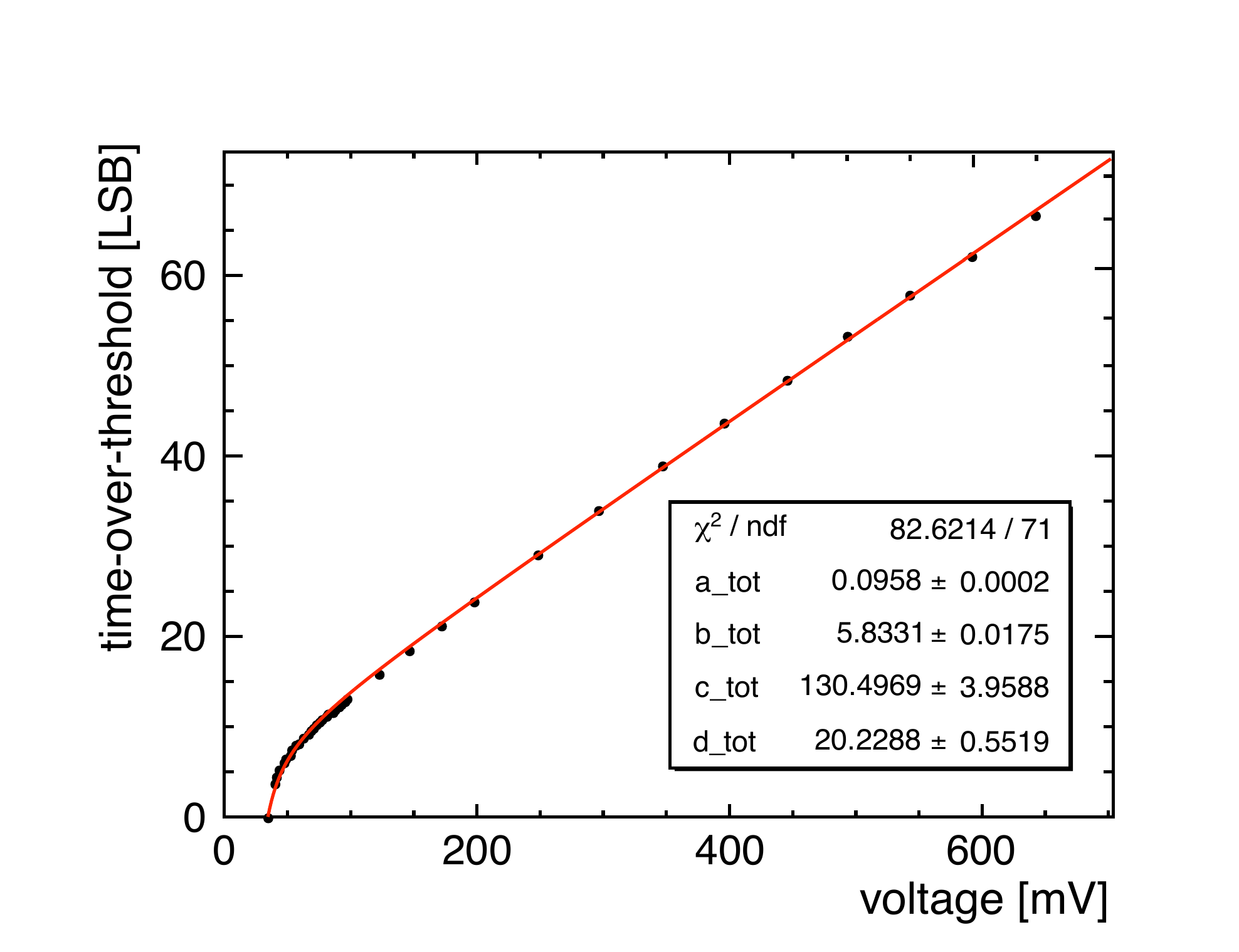}
			\label{fig:tot_example}}
		\caption{Example fits of the timewalk (a) and the time-over-threshold response (b) measured with test pulses for a single pixel.}
		\label{fig:example_fits}
	\end{center}
\end{figure}

The timewalk is calibrated via injection of electrical test pulses of various amplitudes. First, a test pulse is injected directly into the in-pixel digital logic, bypassing the discriminator and thus avoiding timewalk. After a fixed delay, a second test pulse is injected into the analog frontend. The ToA and ToT values of both pulses are recorded and the time difference is calculated. This procedure is repeated 100 times per amplitude and for various amplitudes of the second pulse to sample the dependence of the time difference on the injected amplitude. To avoid crosstalk, only one pixel in a 16 by 16 pixel sub-matrix is pulsed at the same time. The resulting graphs are then fitted with Eq.~\ref{eq:toa} for ToA and Eq.~\ref{eq:tot} for ToT. Two example fits can be seen in Fig.~\ref{fig:example_fits}. The extracted parameters per pixel are used to correct the ToA value based on the measured ToT value. A more detailed description of the injection process and energy calibration can be found in \cite{ref:pitters}.

\begin{figure}[thbp]
	\begin{center}
		\includegraphics[width=0.55\textwidth]{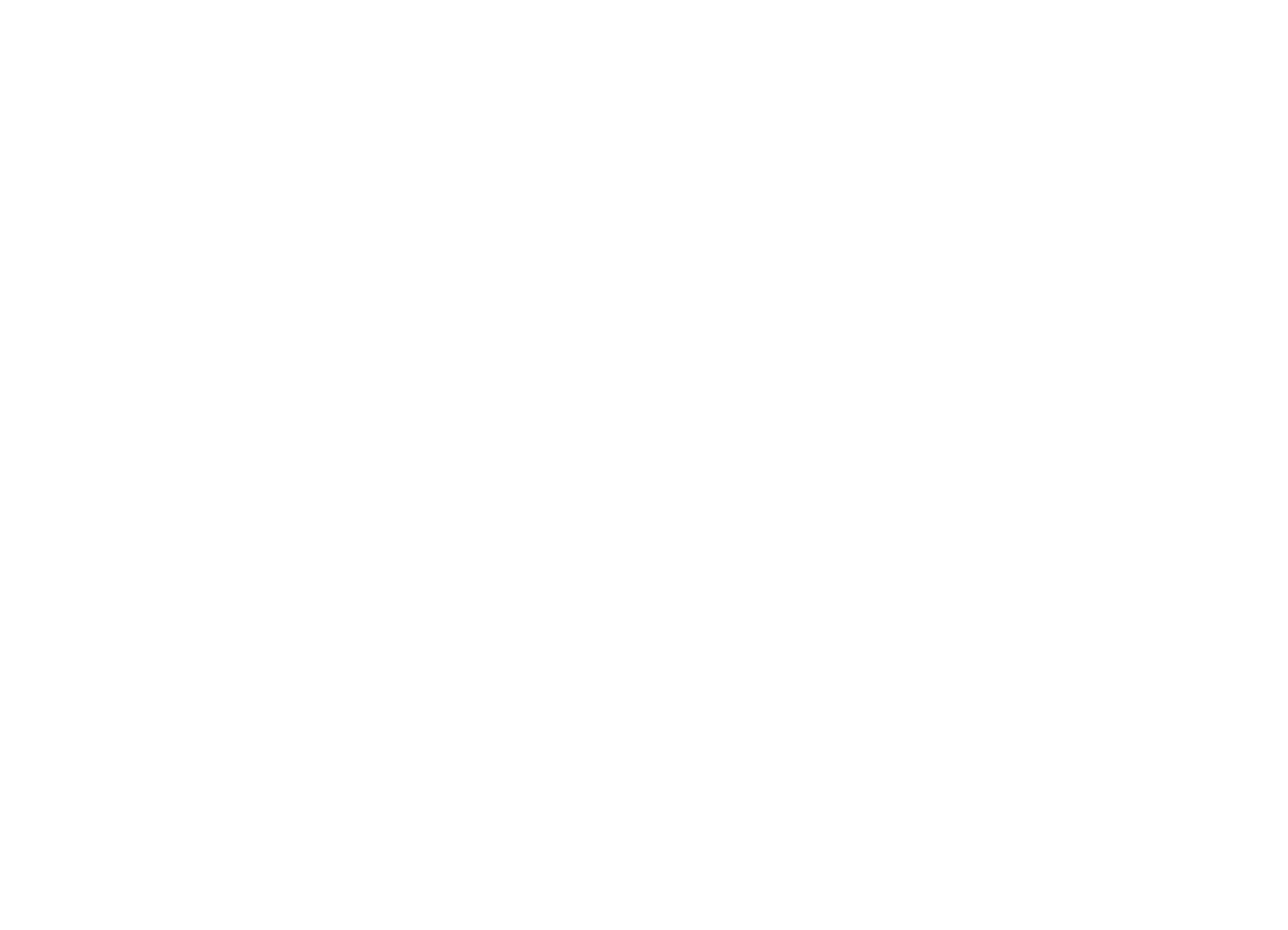}
		\caption{The mean and RMS of the measured time delay averaged over all pixels from 100 injected test pulses per amplitude and pixel.}
		\label{fig:time_vs_contributions}
	\end{center}
\end{figure}

Fig.~\ref{fig:time_vs_contributions} shows the mean and RMS of the measured timewalk averaged over all pixels from 100 injected test pulses per amplitude and pixel. While the mean timewalk can be corrected, the RMS of the timewalk is an irreducible contribution to the time resolution. It decreases with increasing pulse height, reaching a constant value of about 0.42\,ns.
The test pulse injection is not randomised and occurs always at the same clock phase. Therefore, the RMS of the timewalk shown in Fig.~\ref{fig:time_vs_contributions} represents only the jitter on the time measurement due to noise and a finite rise time. An additional contribution to the time resolution of Timepix3 of at least 0.45\,ns from the fine ToA clock period of 1.56\,ns divided by $\sqrt{12}$ is expected.
Taking the root sum of the squares, the best achievable resolution is expected to be 0.62\,ns. Effects from the signal development inside the sensor itself as well as non-uniformities in the clock period and imperfect calibration may degrade this resolution.

\subsection{Delay Calibration}
While the timewalk calibration also includes a delay with the $d_\text{toa}$ parameter (see Eq.~\ref{eq:toa}), it can not detect any variations in the clock itself because the test pulses are injected in phase with the clock and only differences between two pulses are measured. Therefore, beam tests have been conducted in the setup described in Sec.~\ref{sec:setup_testbeam} to calibrate the delay.

\begin{figure}[thbp]
	\begin{center}
    \subfigure[]{
      \includegraphics[width=0.49\textwidth]{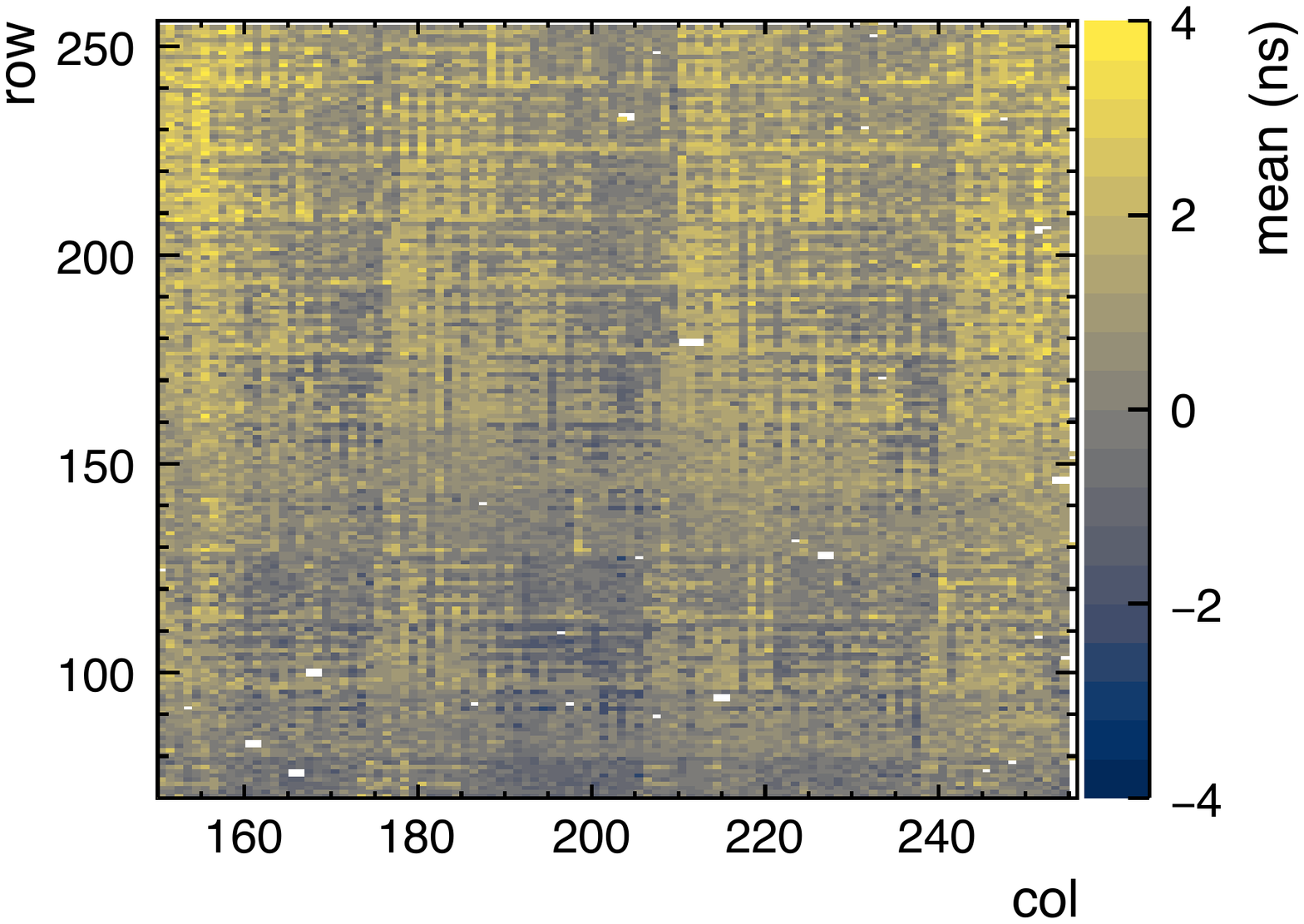}
      \label{fig:meanperpixel}}%
    \subfigure[]{
      \includegraphics[width=0.49\textwidth]{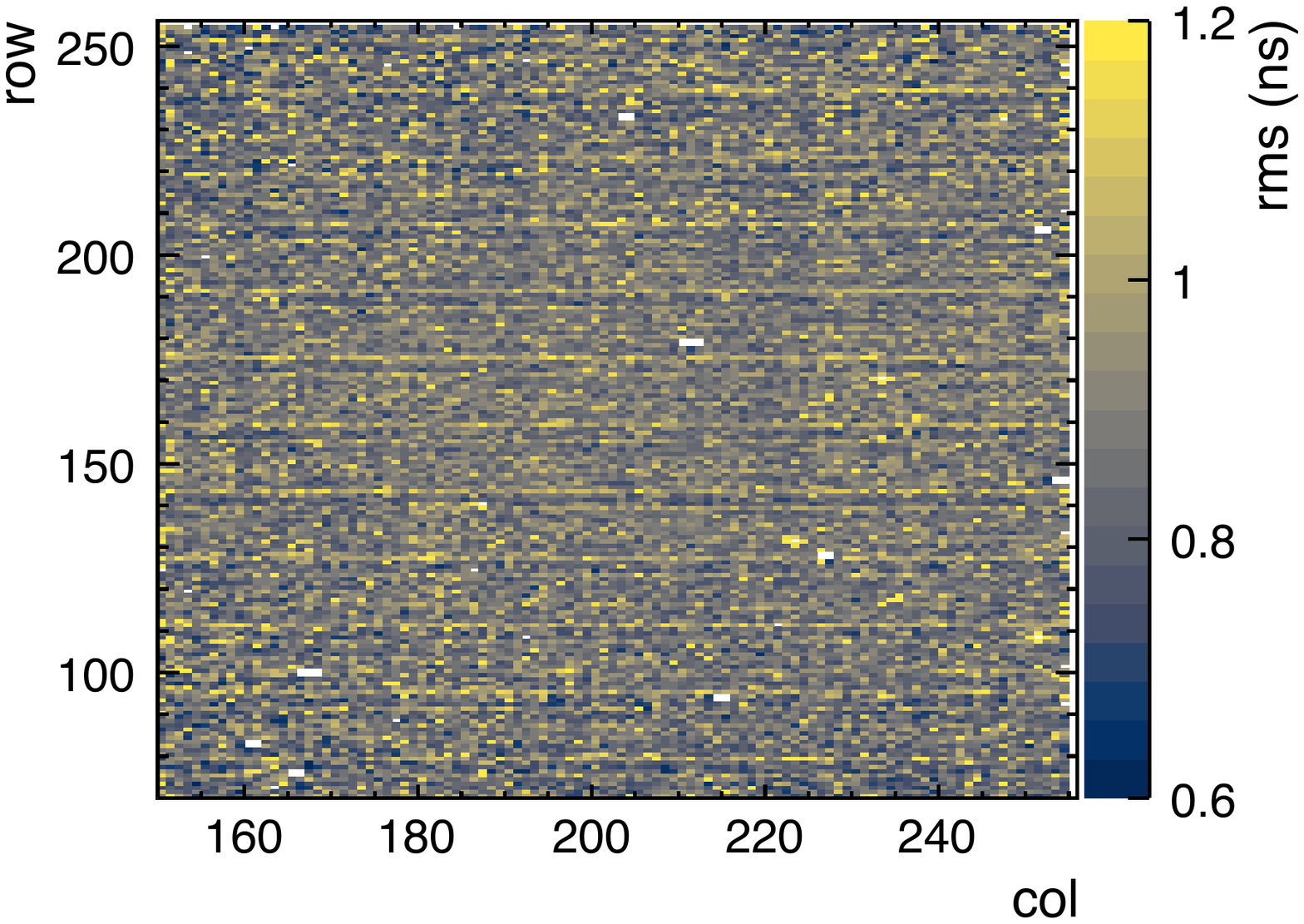}
      \label{fig:rmsperpixel}}
		\caption{(a) Mean of the time difference between DUT and scintillator across the pixel matrix. (b) The RMS of the same distributions.}
    \label{fig:timeperpixel}
	\end{center}
\end{figure}

Fig.~\ref{fig:meanperpixel} shows the mean time difference between scintillator and DUT across the pixel matrix.\footnote{Due to the active-edge characteristic of the sensors, data was only collected in one corner of the assemblies. For this reason, Figs.~\ref{fig:meanperpixel} and \ref{fig:rmsperpixel} do not show the full pixel matrix.} Qualitatively, structures with a period of 16 double columns can be observed and a tendency of later timestamps from higher row numbers. The latter is expected as the clock propagation along one column takes about 1.5\,ns between the top and the bottom row. This value is also observed in data but appears not to be the dominant pattern. The extracted values are used to obtain a pixel by pixel delay calibration. Fig.~\ref{fig:rmsperpixel} shows the RMS for the same distributions. Here, larger values every 16 rows can be correlated to larger noise values that are also observed for these rows.

\section{Results}
\label{sec:results}
\begin{figure}[thbp]
	\begin{center}
		\subfigure[]{
			\includegraphics[width=0.49\textwidth]{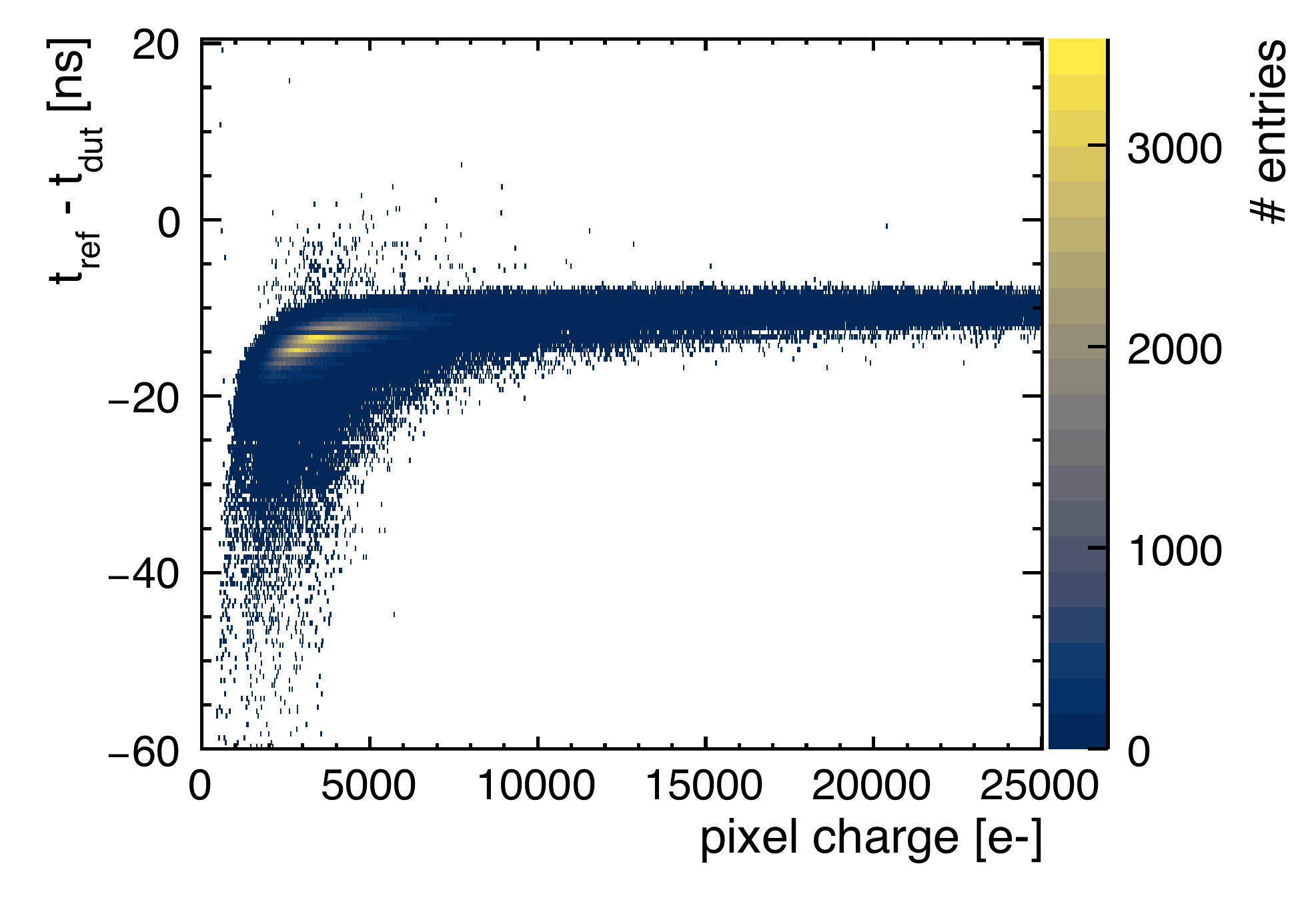}
			\label{fig:timewalk_after_cor:a}}%
		\subfigure[]{
			\includegraphics[width=0.49\textwidth]{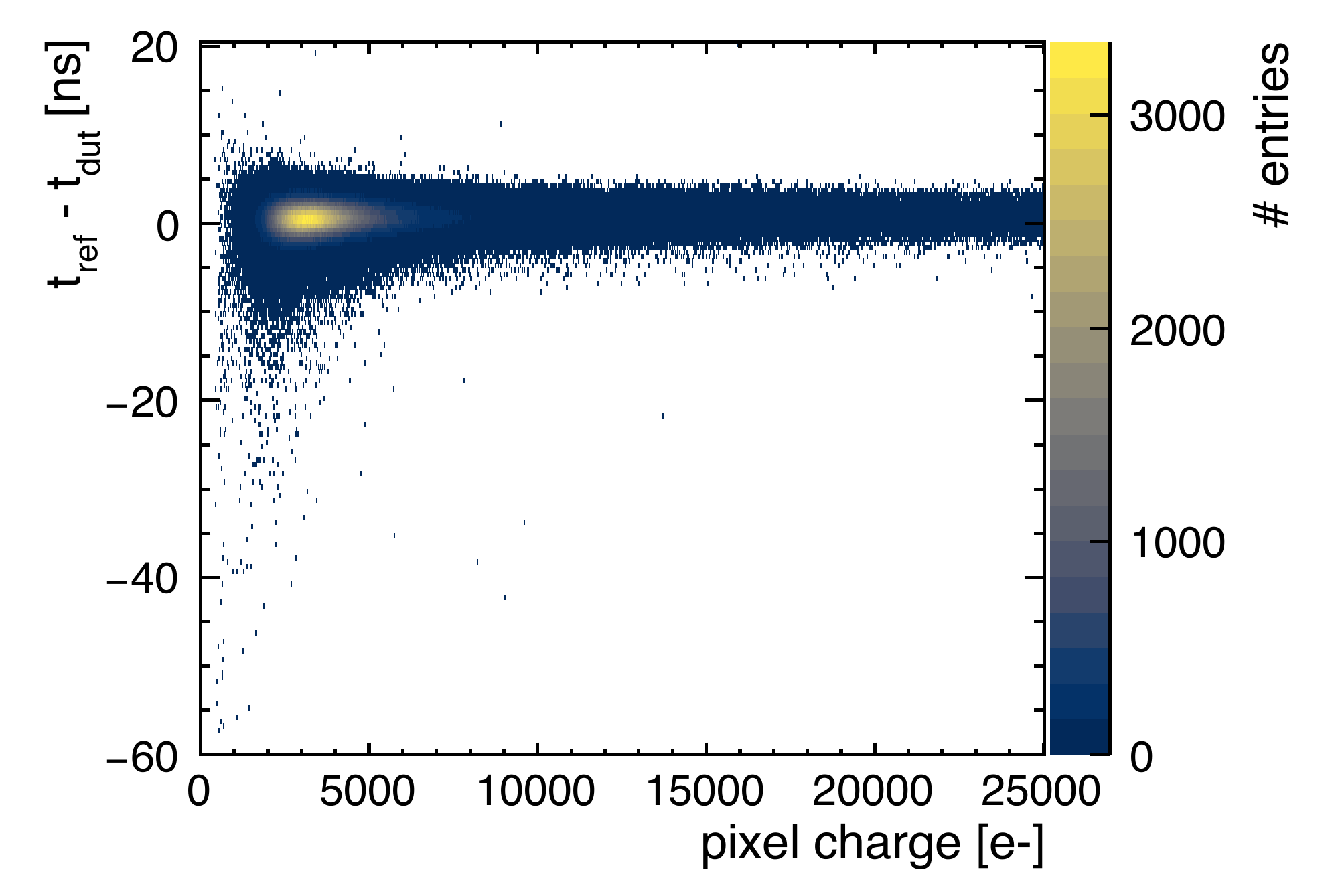}
			\label{fig:timewalk_after_cor:b}}
		\caption{Timewalk (a) before and (b) after the applied calibration corrections for a 50\,\textmu m thick sensor at 5\,V over-depletion.}
		\label{fig:timewalk_after_cor}
	\end{center}
\end{figure}

Fig.~\ref{fig:timewalk_after_cor} shows the extracted time difference as function of the signal amplitude (a) before and (b) after the applied calibration corrections for a 50\,\textmu m thick sensor at 5\,V over-depletion. The remaining asymmetry at the smallest charge deposits after applied corrections can be attributed to specific pixels in which the timewalk calibration process showed poor performance.

\begin{figure}[thbp]
	\begin{center}
		\includegraphics[width=0.75\textwidth]{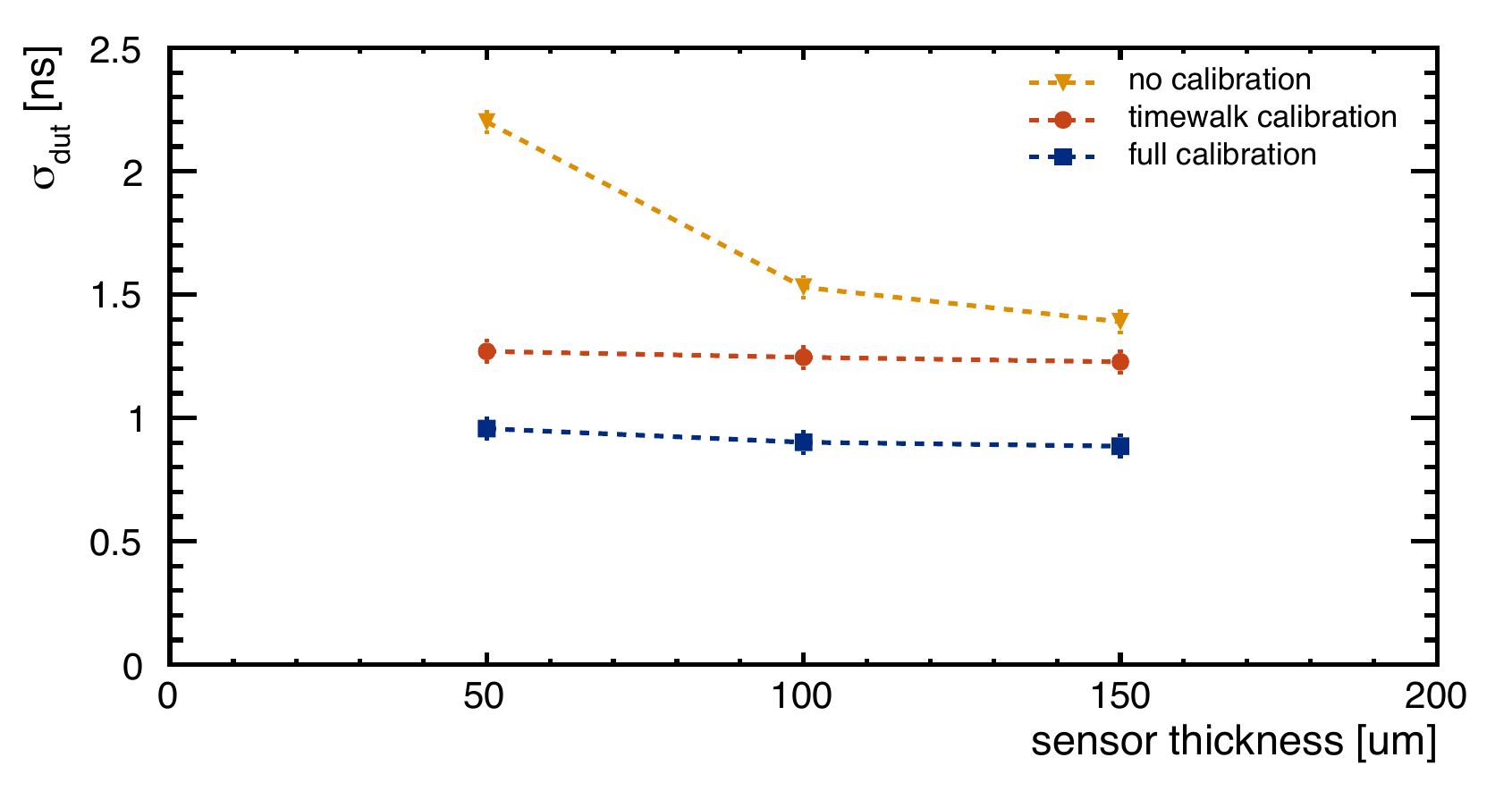}
		\caption{Time resolution with and without timewalk and delay corrections for 50\,\textmu m, 100\,\textmu m and 150\,\textmu m thick sensors at 5\,V over full depletion. The reference timestamp is taken with an organic scintillating read out by a PMT. Its resolution of 0.3\,ns is deconvoluted.}
		\label{fig:time_vs_thickness}
	\end{center}
\end{figure}

Fig.~\ref{fig:time_vs_thickness} shows the time resolution for sensor thicknesses of 50\,\textmu m, 100\,\textmu m and 150\,\textmu m sensor, each operated at 5\,V over-depletion. A significant improvement in the resolution can be seen due to the timewalk correction for the 50\,\textmu m sensor. The improvement is less pronounced for thicker sensors due to the larger average energy deposits. After timewalk and delay corrections, all sensors yield approximately the same resolution at 5\,V over-depletion.

\begin{figure}[thbp]
	\begin{center}
		\includegraphics[width=0.75\textwidth]{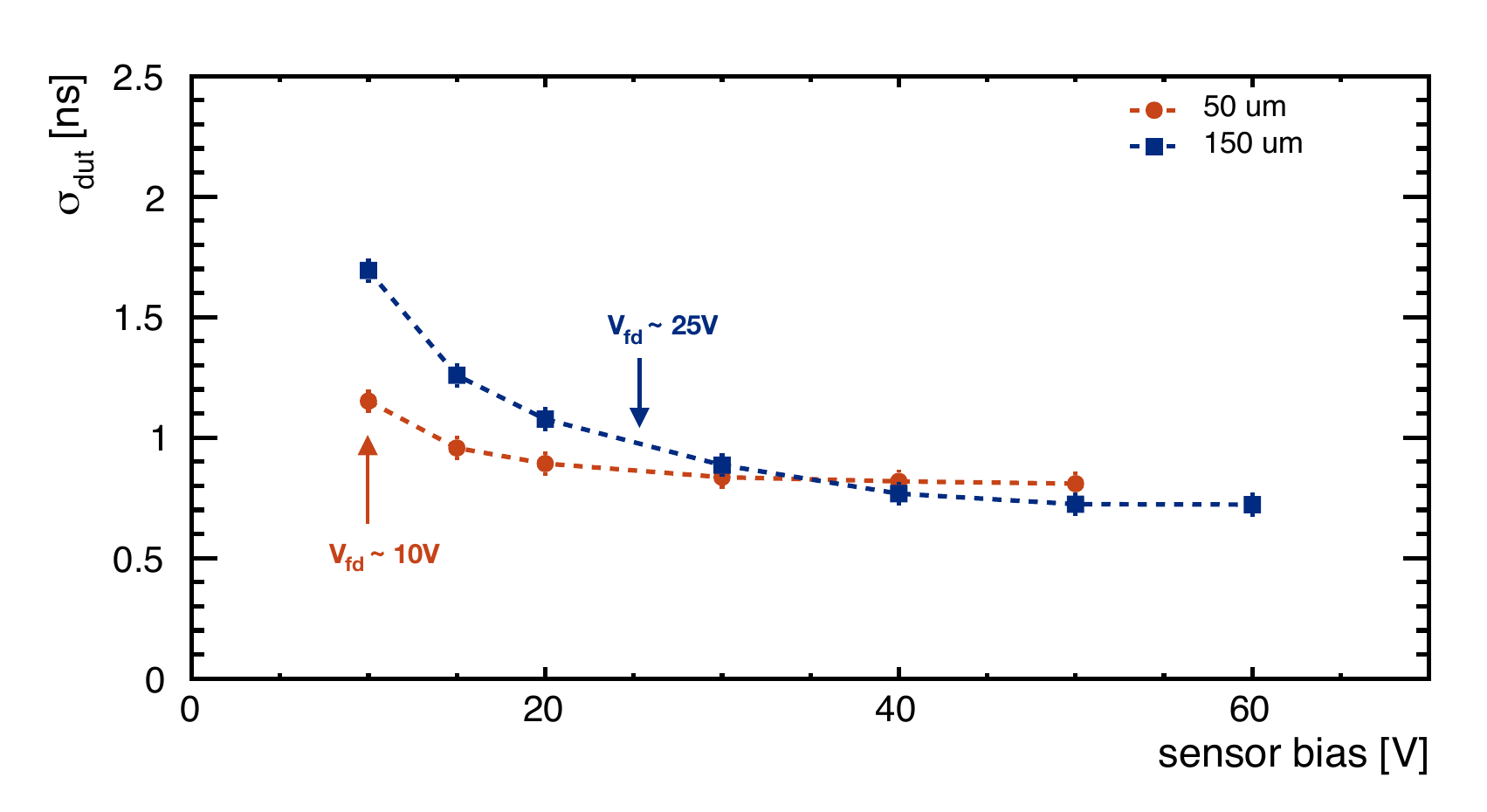}
		\caption{Time resolution with timewalk and delay corrections at different bias voltages and for 50\,\textmu m and 150\,\textmu m thick sensors. The full depletion voltages V$_\text{fd}$ are indicated. An improvement beyond full depletion can be seen due to stronger electric fields.}
		\label{fig:time_vs_bias}
	\end{center}
\end{figure}

The electric field also influences the time resolution. Fig.~\ref{fig:time_vs_bias} shows the obtained time resolution for different bias voltages for the 50\,\textmu m and 150\,\textmu m sensor which are fully depleted at around 10\,V and 25\,V respectively. The resolution however improves beyond that point due to the stronger electric field in the sensor at higher bias voltages.
This in turn increases the drift velocity which results in a faster and more uniform signal creation on the CSA input and, owing to reduced diffusion, a trend towards larger pixel charges. A slightly better resolution can be observed for the 150\,\textmu m sensor at the highest tested bias voltage. This is attributed to the larger average charge deposition in the 150\,\textmu m sensor. The best value for the time resolution reached in this study is 0.72\,$\pm$\,0.04\,ns for a 150\,\textmu m thick sensor at 35\,V over-depletion. This is close to the expected limit found in Sec.~\ref{sec:tw_calibration}.

\begin{figure}[thbp]
	\begin{center}
		\subfigure[]{
			\includegraphics[width=0.49\textwidth]{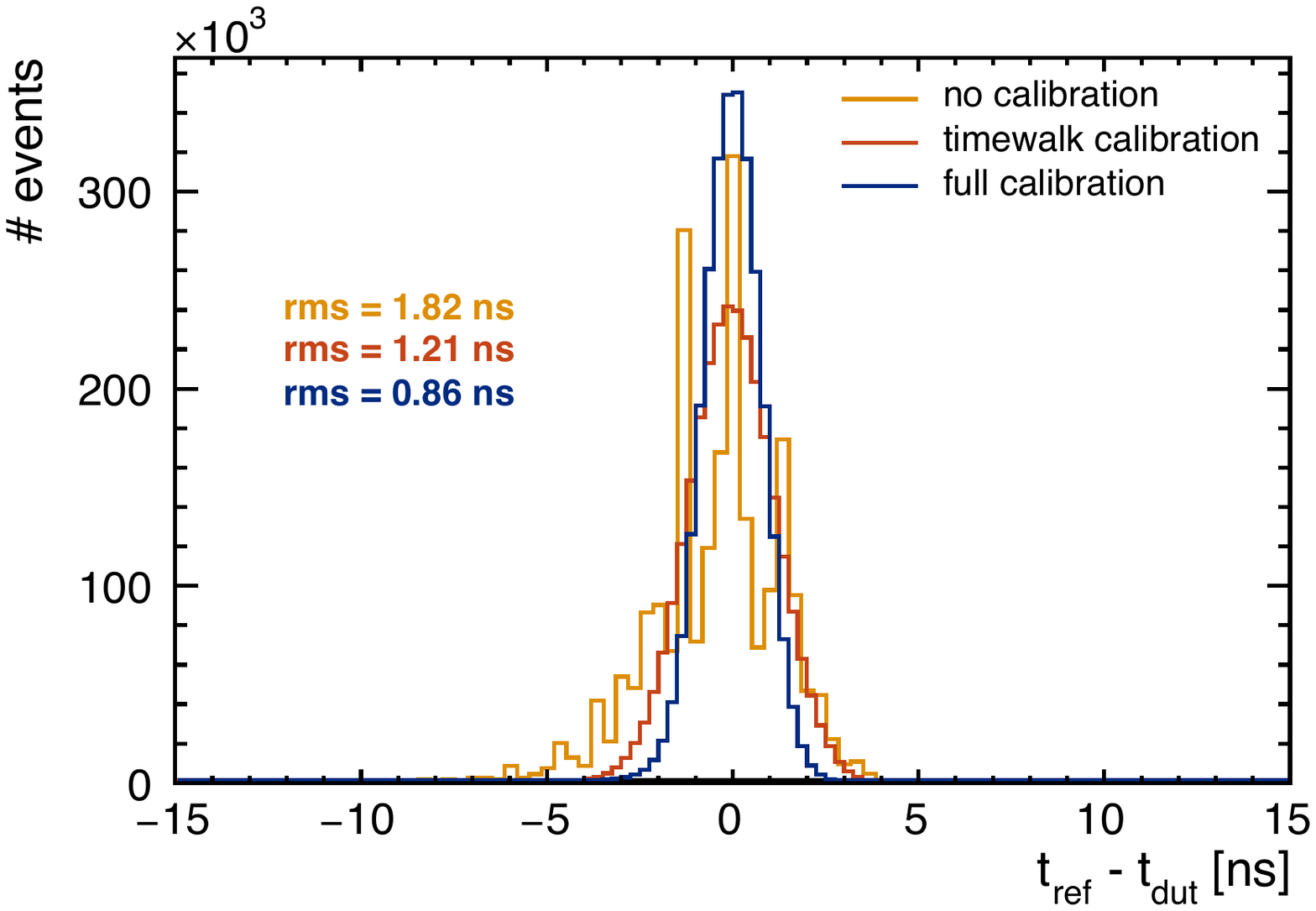}
			\label{fig:example_histos1}}%
		\subfigure[]{
			\includegraphics[width=0.49\textwidth]{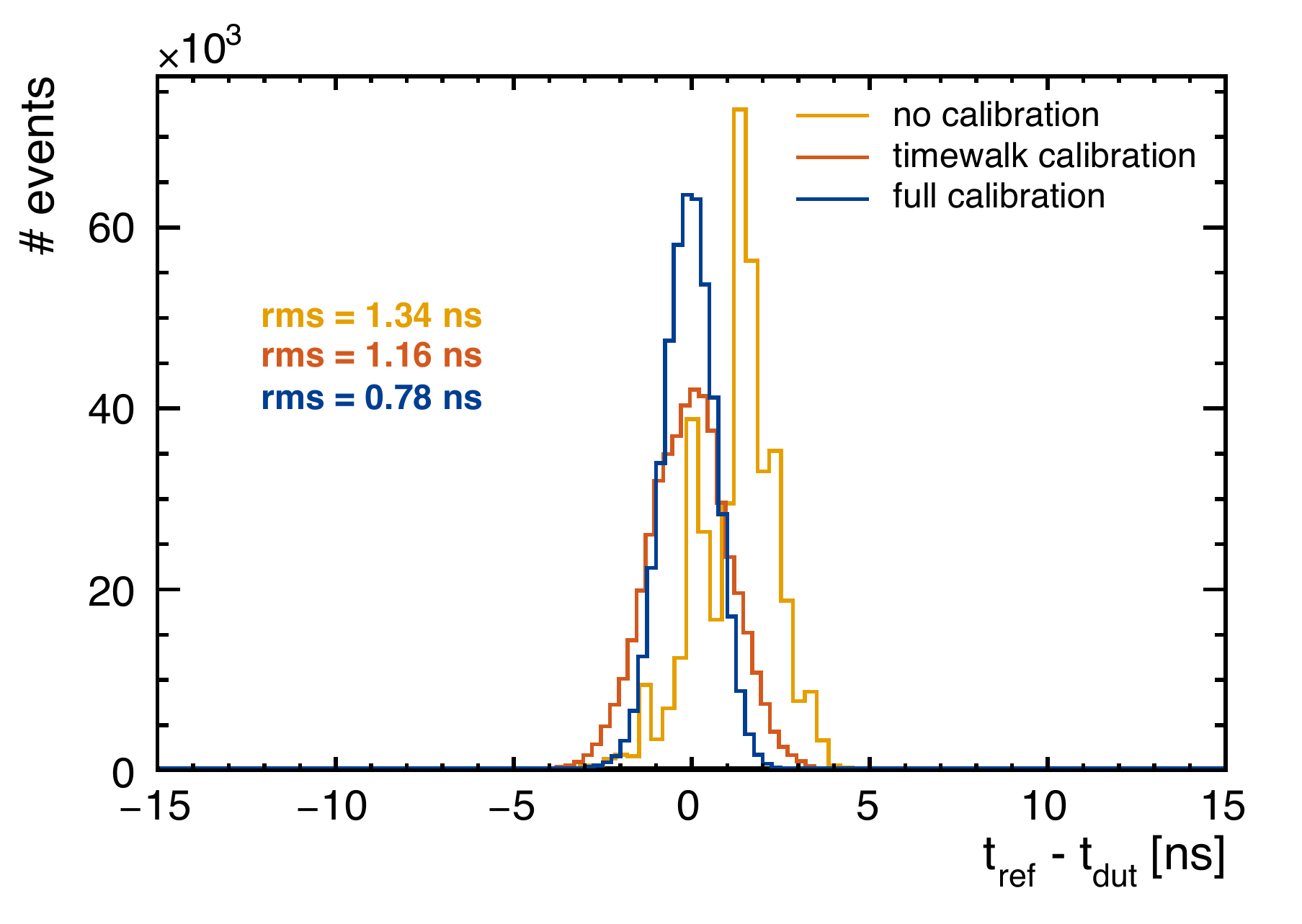}
			\label{fig:example_histos2}}
		\caption{Example histograms for (a) 50\,\textmu m at 50\,V and (b) 150\,\textmu m at 60\,V at different calibration stages. From the RMS values, 0.30\,ns are deconvoluted to obtain the time resolution.}
		\label{fig:time_vs_bias_examples}
	\end{center}
\end{figure}

Fig.~\ref{fig:time_vs_bias_examples} shows the distributions of t$_\text{ref}$\,-\,t$_\text{dut}$ for the highest bias voltages of the two sensor thicknesses at the various calibration stages. The uncalibrated histogram is highly skewed due to timewalk and shows multiple peaks as pixel by pixel variations are not yet compensated. After applying both calibration steps, the multiple peak structures disappear and the histograms become Gaussian.

\section{Summary and Conclusions}
\label{sec:summary}
In this work, a method for pixel-by-pixel calibration of the Timepix3 ASIC has been presented. The method uses a combination of electrical test pulses and beam data and requires only about 100 hits per pixel from minimum ionising particles. Timepix3 assemblies with thin silicon sensors of 50\,\textmu m to 150\,\textmu m have been assessed in particle beams. An improvement in the time resolution beyond full depletion is observed. For a 150\,\textmu m thick sensors and full calibration, time resolutions down to 0.72\,ns are achieved. This is close to the expected limit of Timepix3. From these results it can be deduced that the contributions to the time resolution from the signal development in thin planar silicon sensors are well below 0.72\,ns, providing the timewalk is adequately corrected. This is significantly better than required for the CLIC vertex detector.

\acknowledgments
The authors thank Fernando Duarte Ramos (CERN) for his support with the mechanical integration of the tested devices in the telescope system. The help from the staff operating the CERN SPS and the North Area test facilities is gratefully acknowledged. This project has received funding from the European Union's Horizon 2020 Research and Innovation programme under Grant Agreement no. 654168 and from the Austrian Doctoral Student Programme at CERN.


\end{document}